\documentclass[12pt,elsart,epsfig]{article}     % Paper
\usepackage{epsfig}
\begin{document}

\begin {center}
{\Large \bf Combined analysis of meson channels with $I=1$, $C = -1$
from 1940 to 2410 MeV}
\vskip 5mm
{A.V. Anisovich$^d$, C.A. Baker$^a$, C.J. Batty$^a$, D.V. Bugg$^c$, L.Montanet$^b$,  
V.A. Nikonov$^d$, A.V. Sarantsev$^d$, V.V. Sarantsev$^d$, 
B.S.~Zou$^{c}$ \footnote{Now at IHEP, Beijing 100039, China} \\
{\normalsize $^a$ \it Rutherford Appleton Laboratory, Chilton, Didcot OX11 0QX,UK}\\
{\normalsize $^b$ \it CERN, CH-1211 Geneva, Switzerland }\\
{\normalsize $^c$ \it Queen Mary and Westfield College, London E1\,4NS, UK}\\
{\normalsize $^d$ \it PNPI, Gatchina, St. Petersburg district, 188350, Russia}\\ 
[3mm]}
\end {center}

\begin{abstract}
New Crystal Barrel data are reported for $\bar pp \to \omega \pi ^0$
and $\bar pp \to \omega \eta \pi ^0$ with $\omega$ decaying to
$\pi ^+\pi ^-\pi ^0$.
The shapes of angular distributions agree well with those
for data where $\omega \to \pi ^0 \gamma$;
this is a valuable cross-check on systematic errors.
The new data provide good measurements of vector and tensor
polarisations $P_y$, $T _{20}$, $T_{21}$ and $T_{22}$ of the $\omega$.
These lead to significant improvements in parameters of several
resonances reported earlier.
New values  of masses and widths (in Mev) are:
$J^{PC} = 5^{--}$ ($2300 \pm 45$, $260 \pm 75$),
$J^{PC} = 3^{--}$ ($2260 \pm 20$, $160 \pm 25$),
$J^{PC} = 1^{+-}$ ($2240 \pm 35$, $320 \pm 85$),
and $J^{PC} = 1^{--}$ ($2110 \pm 35$, $230 \pm 50$).
A remarkable feature of the data is that vector polarisation $P_y$ is
close to zero everywhere.
It follows that all interfering amplitudes have relative phases close to
0 or 180$^{\circ  }$.
Tensor polarisations are large.
\end{abstract}

Earlier publications have reported Crystal Barrel data for $\bar pp \to
\omega \pi ^0$ [1] and $\bar pp \to \omega \eta \pi ^0$ [2] in
all-neutral final states; there, $\omega$ decays to $\pi ^0
\gamma$.
Data for $\bar pp \to \omega \pi ^0$ have also been reported by Peters
[3].
A weakness of all these data is that much of the information concerning
$\omega $ polarisation is carried away by the photon, whose
polarisation is not measured.
Here we report data where $\omega$ decays to $\pi ^+\pi ^- \pi ^0$.

 The matrix element for this decay is relativistically $\epsilon
_{\alpha \beta \gamma \delta} p^{\beta }p^{\gamma } p^{\delta }$, where
$p$ are 4-momenta of decay pions. Non-relativistically this matrix
element in the rest frame of the $\omega $ becomes $p _i \wedge p_j$
where $p_i$ are 3-momenta of any two pions, e.g. $\pi ^+$ and $\pi ^-$.
The polarisation vector of the $\omega$ is therefore described by the
normal $\vec n$ to its decay plane.
A measurement of the decay plane provides complete information
on the polarisation of the $\omega$.
This information improves considerably the determination of masses and
widths of several resonances reported in Refs. [1] and [2].
It leads to an almost complete spectrum of resonances consistent with
$q\bar q$ states expected in the available mass region 1960--2410 MeV.

The data were collected at LEAR at 6 momenta from 600 to 1940 MeV/c
using the Crystal Barrel detector [4].
Photons are detected in a barrel of 1380 CsI cystals covering 98\% of
the solid angle.
Charged pions are detected in a JET drift chamber which
surrounds the 4.4 cm liquid hydrogen target.
This chamber is cylindrical with its axis parallel to the beam.
It contains 24 layers measuring momenta in the solenoidal
field of 1.5T.
The JET chamber provides full coverage only over the lab angular range
$|\cos \theta | \le 0.71$.
Particles are thrown forwards by the Lorentz boost due to beam momentum.
In consequence, $\omega$ are detected efficiently only in the
backward hemisphere in the centre of mass.
This is adequate, since conservation of C-parity demands that the
production angular distribution is symmetric forwards and backwards
with respect to the beam.

\begin{table} [htp]
\begin{center}
\caption {Numbers of selected events (including backgrounds).}
\begin{tabular}{ccc}
\hline
Beam Momentum & $\omega \pi ^0$ & $\omega \eta \pi ^0$ \\
  (MeV/c)& & \\ \hline
600 & 3590 & 719 \\
900 & 17978 & 4171 \\
1200 & 7912 & 2577  \\
1525 & 2651 &  998 \\
1642 & 4787 & 2464  \\
1940 & 1476 &  737 \\\hline
\end{tabular}
\end{center}
\end{table}
A preliminary selection of both $\omega \pi ^0$ and $\omega \eta
\pi ^0$ events requires that both charged particles are produced with
$|\cos \theta | \le 0.65$, in order to avoid edge-effects in the JET
chamber.
At least 11 digitisations are demanded, including hits in at least one
of the first 3 layers and at least one in the last three.
Events are required to have exactly the right number of photons:
4 for $\omega \pi ^0$ and 6 for $\omega \eta \pi ^0$.
In the final selection of $\omega
\pi ^0$ and $\omega \eta \pi ^0$, the following cuts are applied: (a)
confidence level $(CL)$ for the signal channel $> 10\%$; (b) $CL$ of
the signal channel higher than that for any other channel (which all
have much lower branching fractions); (c) as a minor refinement to
check that the $\omega$ is well reconstructed, it is required that
$CL(\omega \pi ^0) > 0.8 \times CL(\pi ^+\pi ^- \pi ^0 \pi ^0)$ and
$CL(\omega \pi ^0) > 0.5 \times CL(\pi ^+\pi ^- 4\gamma )$;
corresponding cuts are applied to the selection of $\omega \eta \pi
^0$. Table 1 summarises the number of selected events.

Figs. 1(a) and (b) show illustrative plots at one momentum of the
$\pi ^+\pi ^-\pi ^0$ mass distribution, before the kinematic fit to
the $\omega$.
One sees a clear $\omega$ signal above a rising background.
Both $\pi ^0$ combinations are included.
The few events where both combinations lie in the mass range
760--804 MeV are rejected.
For the final selection of events, a kinematic fit is applied to
$\omega \pi ^0$ or $\omega \eta \pi ^0$, constraining the $\omega$
mass to the value 781.95 MeV.
This narrows somewhat the mass range over which $\omega$ are selected
and improves the signal/background ratio.
In determining the efficiency of reconstruction, we break the $\omega$
peak of Figs. 1(a) and (b) into bins 3 MeV wide and track the
efficiency with which events in each bin pass the final kinematic fit.
We also allow for the contribution to background from `wrong'
combinations of the spectator $\pi ^0$ with $\pi ^+\pi ^-$. In this
way, the background under the $\omega$ peak is estimated to vary with
beam momentum from 14 to 20\% for $\omega \pi ^0$. It is 26--36\% for
$\omega \eta \pi ^0$.
%Fig. 1
\begin{center} 
\begin{figure}
\centerline{\hspace{0.2cm}\epsfig{file=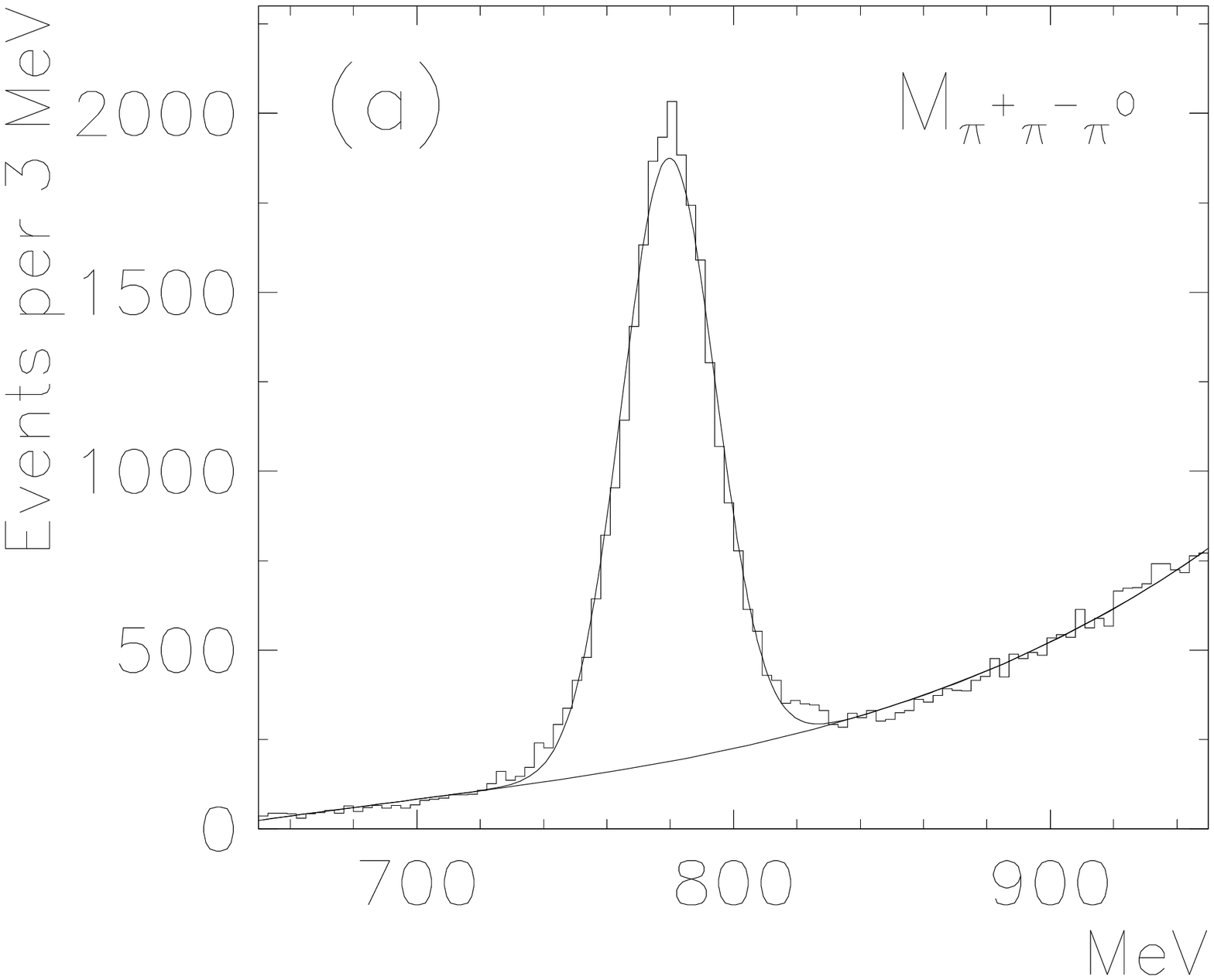,width=5.1cm}
                    \epsfig{file=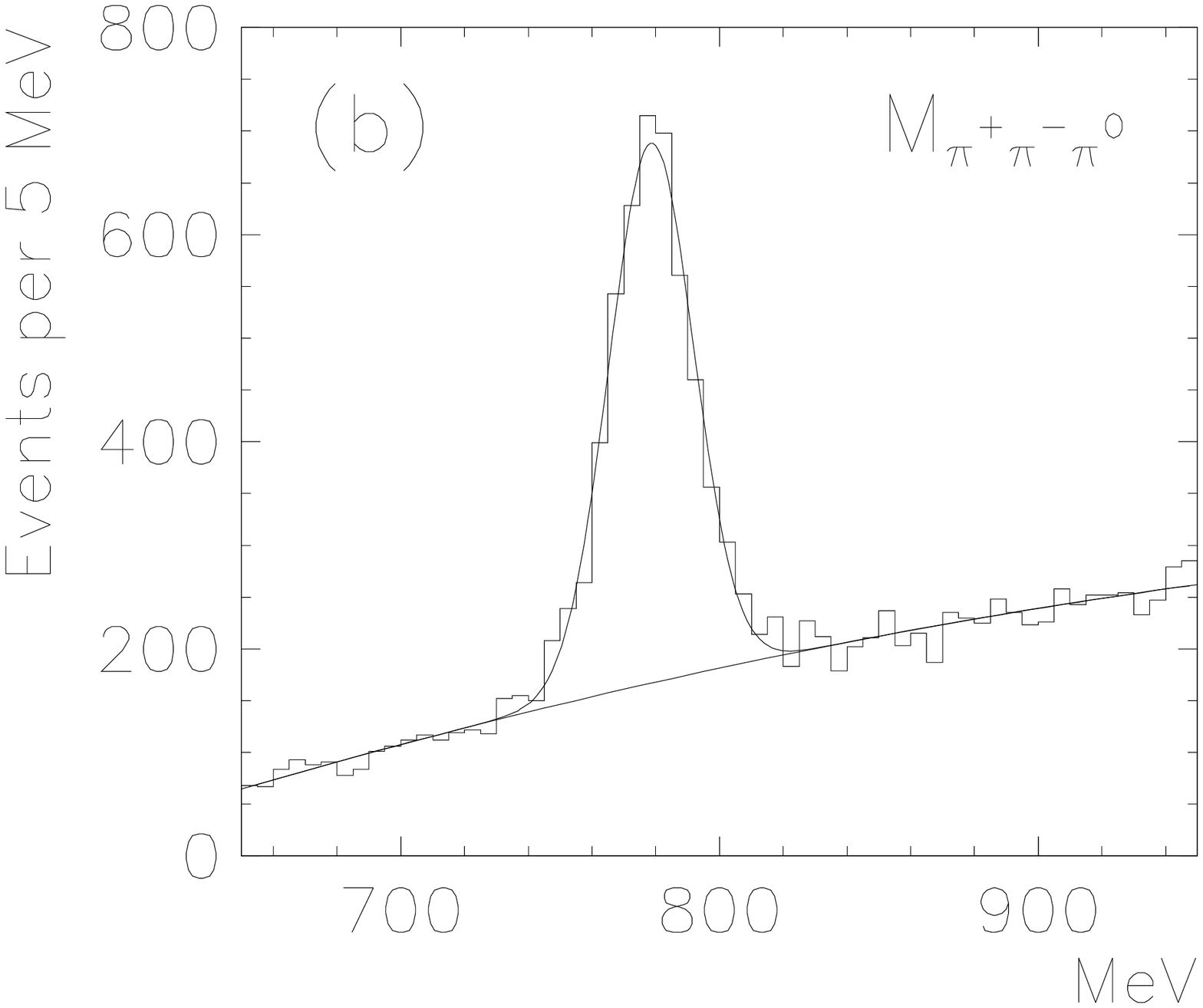,width=5.1cm}}
\vskip -5.14cm
\centerline{\hspace{0.2cm}\epsfig{file=F1A_WPI.PS,width=5.1cm}
             \epsfig{file=F1B_WPI.PS,width=5.1cm}}
\vspace{-1.1cm}
\centerline{\hspace{0.3cm}\epsfig{file=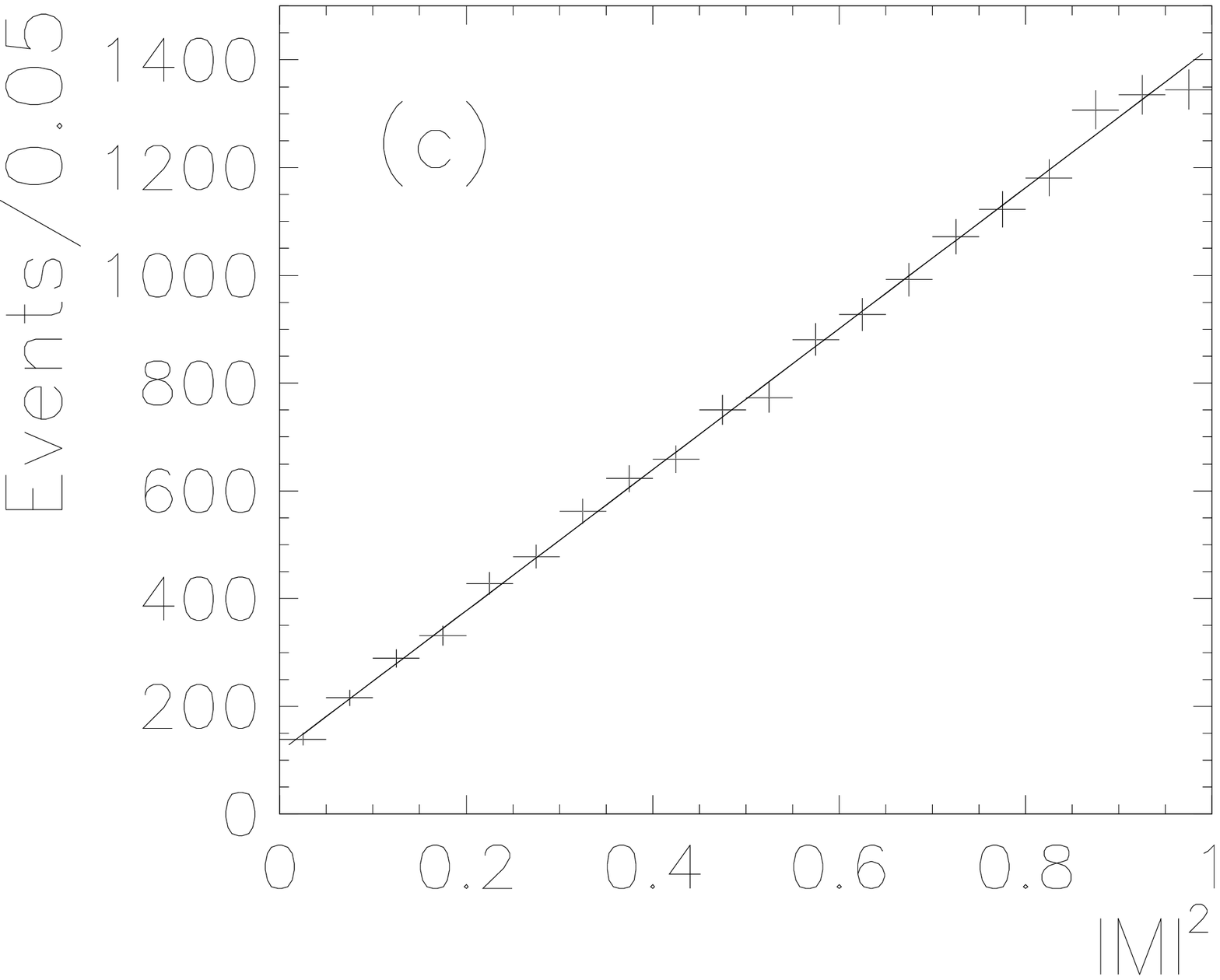,width=5.2cm}
            \epsfig{file=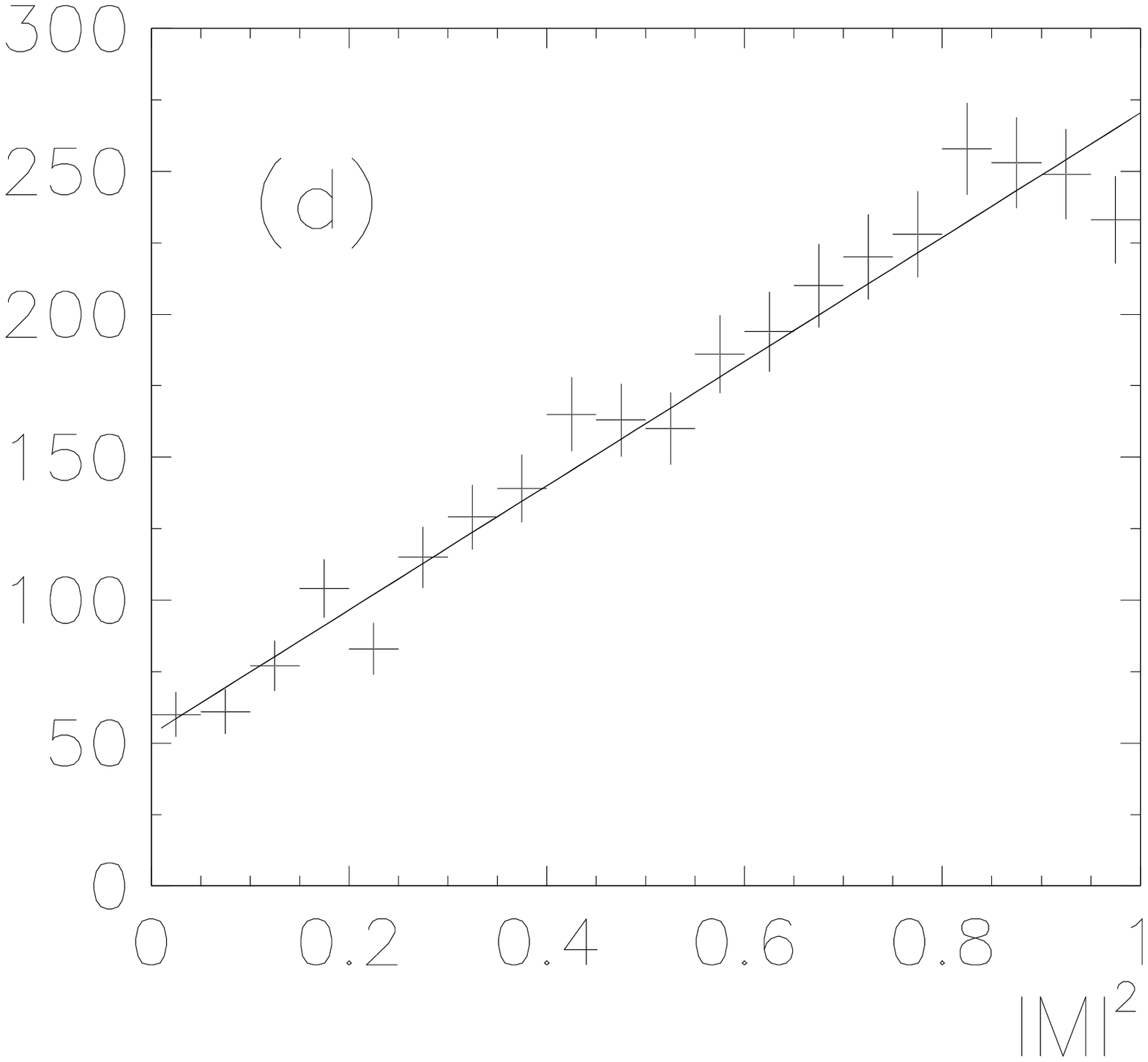,width=5.2cm}}
\vskip -5.24cm
\centerline{\hspace{0.3cm}\epsfig{file=F1C_WPI.PS,width=5.2cm}
            \epsfig{file=F1D_WPI.PS,width=5.2cm}}
\vspace{-1.5cm}
\centerline{\epsfig{file=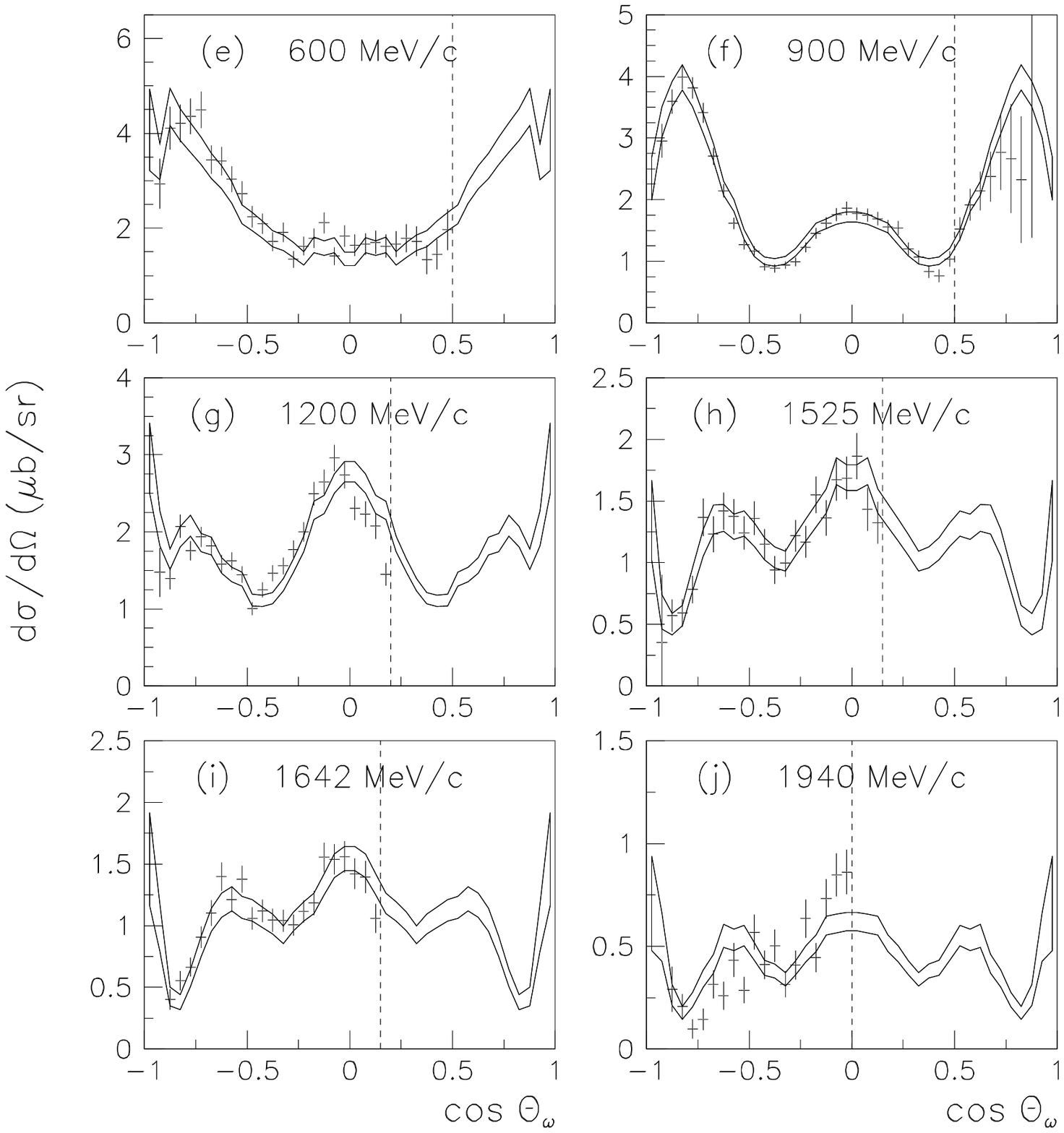,width=13cm}}
\vskip -130.25mm
\centerline{\epsfig{file=F1EJ_WPI.PS,width=13cm}}
\vspace{-0.5cm}
\caption{$M(\pi ^+\pi ^- \pi ^0)$ from the preliminary data selection
for (a) $\omega \pi ^0$ data, (b) $\omega \eta \pi ^0$ at 900 MeV/c;
the number of events v. the square of the matrix element for
$\omega$ decay in (c) $\omega \pi ^0$, (d) $\omega \eta \pi ^0$;
(e)-(j): curves show the error corridor for $\omega \pi ^0$ differential
cross sections where $\omega$ decays to $\pi ^0 \gamma$;
they are corrected for angular acceptance; points with
errors show new results for $\omega$ decays to $\pi ^+\pi ^-\pi ^0$;
the vertical dashed lines show the cut-off which has been used.}
\end{figure}
\end{center}

The background may be estimated in a second way.
It is well known that decays of the $\omega$ are enhanced near
the edge of its Dalitz plot because of the matrix element for
decay.
Figs. 1(c) and (d) show plots of the number of events against the
square of this matrix element.
One sees straight lines with intercepts which provide another
estimate of the background; it agrees with the first within errors.
This background is included into the partial wave analysis described
below,
using Monte Carlo events which pass the data selection;
they are generated according to $\pi ^+\pi ^- \pi ^0 \pi ^0$ or
$\pi ^+\pi ^-\eta \pi ^0 \pi ^0$ phase space.

We now compare angular distributions for present $\omega \pi ^0$ data
with those determined from all-neutral data where $\omega \to \pi
^0\gamma$.
Both are corrected for acceptance.
Figs. 1(e)-(j) show error corridors through the latter data.
Points with errors are superposed from present data.
The absolute efficiency for tracking charged particles
has a significant uncertainty;
it is sensitive to precise cuts on the number
of layers and the $\chi ^2$ for the fit to to a helix.
Therefore, the absolute scale for present data is normalised to that for
all-neutral data.

For the backward hemisphere for $\omega $ production,
shapes of angular distributions in Figs. 1(e)--(j) agree well
between present data and the earlier publication.
(A possible exception is at 1940 MeV/c, but statistics are lowest
there). This is a valuable cross-check on experimental techniques,
particularly at 900 MeV/c, where statistics are highest.
Systematics of calibrations and data selection are completely different
between all-neutral data and the present data where $\omega \to
\pi ^+\pi ^-\pi ^0$.
In particular, the treatment of the vertex is quite different.
Here, the vertex is determined by the intersection of the two charged
tracks.
For neutral data, the vertex is instead assumed to be at the centre of
the target.
This distorts the kinematic fit slightly,
but the correction is only $\sim 1\%$ to the differential
cross section for all-neutral data [1];
we consider it safer to use this procedure than
fitting the vertex freely, since that alternative leads to a strong
variation of acceptance with $\cos \theta _{\omega }$.
The comparisons  in Figs. 1(e)-(j) vindicate our treatment of
all-neutral data.

In the forward hemisphere, we reject events where the acceptance for
the $\omega$ drops rapidly.
This requires a cut $\cos \theta _{\omega}< 0.5$ at 600 and 900 MeV/c,
$\cos \theta _{\omega}< 0.2$ at 1200 MeV/c, and
$\cos \theta _{\omega}< 0.15$ at higher momenta.
Outside this cut, one sees on Fig. 1(f) an increase in errors and a
possible small asymmetry with respect to the backward hemisphere.
It is safer to discard 10\% of events outside this cut than risk using
events where the acceptance is varying rapidly.
At 1940 MeV/c, a cut of $\cos \theta _{\omega} < 0 $ is used for
differential cross sections, but polarisations are determined to $\cos
\theta _{\omega } = 0.15$, since they depend only on asymmetries.

We now discuss results for the polarisation of the $\omega$.
The formalism for partial wave analysis of $\bar pp \to \omega \pi$
is identical to that for $pp \to d\pi ^+$, given by Weddigen [5] and
Foroughi [6].
Let the spin vector of the $\omega$ be $S$.
Foroughi shows that the vector polarisation of the $\omega$ is non-zero
only along the normal $y$ to the production plane.
There are three tensor polarisations.
We follow the conventional definitions [7]
\begin {eqnarray}
T_{2, \pm 2} &=& \frac {\sqrt {3}}{2}(S_x \pm iS_y)^2 \\
T_{2, \pm 1} &=& \mp \frac {\sqrt {3}}{2}[(S_x \pm iS_y)S_z
+ S_z(S_x \pm iS_y)]  \\
T_{2,0} &=& \sqrt{\frac 12}(3S_z^2 - 2).
\end {eqnarray}
From the formulae of Foroughi, it is readily shown that  imaginary
parts of $T_{2, \pm 2}$ and $T_{2, \pm 1}$ are zero;
also the components of vector polarisation in the plane of scattering
are zero.
We have verified that the present data are everywhere consistent within
errors with those predictions.

Suppose the normal $\vec n$ to the $\omega$ decay plane in its rest
frame is described by polar angle $\alpha$ (with respect to the
beam direction) and azimuthal angle $\beta$.
It may be shown that
\begin {eqnarray}
P_y &\propto & \sin 2\theta _{\omega }\sin \alpha \sin \beta f_1(\cos
^2 \theta _{\omega}) \\
Re T_{2,2} &\propto & \sin ^2\alpha \cos 2\beta f_2(\cos ^2
 \theta _{\omega }) \\
Re T_{2,1} &\propto & \sin 2\theta _{\omega }\sin 2\alpha \cos \beta
f_3(\cos ^2\theta _{\omega }) \\
(T_{20} + 1/\sqrt {2}) &\propto & \sin 2\alpha
\cos 2\beta f_4(\cos  ^2\theta _{\omega}).
\end {eqnarray}
Here $f_i$ are
polynomials in $\cos ^2 \theta _{\omega }$, where $\theta _{\omega }$ is
the centre of mass angle at which the $\omega$ is produced.
So $P_y$, $T_{20}$, $Re~T_{21}$ and $Re~T_{22}$ are determined from
their distinctive dependence on $\alpha$ and $\beta$.
Corrections are applied for detector asymmetries in the
measurement of $T_{20}$, $T_{21}$ and $T_{22}$.

%Fig. 2
\begin{center}
\begin{figure}
\vskip -13mm
\centerline{\epsfig{file=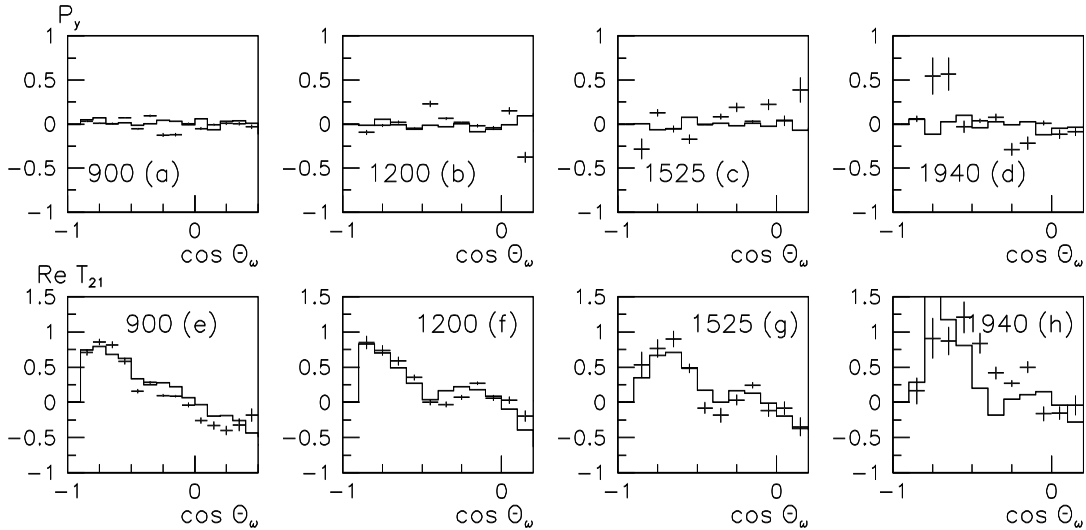,width=15cm}}
\vskip -150.25mm
\centerline{\epsfig{file=f21_wpi.ps,width=15cm}}
\vskip -78mm
\caption{ Vector polarisation $P_y$ at (a) 900, (b) 1200, (c) 1525 and
(d) 1940 MeV/c
compared with the partial wave fit (histogram); (e)--(h) $Re~T_{21}$
at the same momenta. }
\end{figure}
\end{center}
Fig. 2 shows values of $P_y$ and $Re~T_{21}$ at four momenta;
Fig. 3 shows values of $Re~T_{22}$ and $T_{20}$.
Tensor polarisations are large. Values of
$T_{21}$ should lie in the range $-2/\sqrt {3}$ to $+2/\sqrt {3}$,
$T_{22}$ in the range $-\sqrt {3}/2$ to $+\sqrt {3}/2$,
and $T_{20}$ in the range $-\sqrt {2} $ to $1/\sqrt {2}$.
Some experimental values and fitted values stray just outside these
limits; this is because of statistical fluctuations in data
(points with errors)
and because of statistical fluctuations in the Monte Carlo events used
to generate histograms from the maximum likelihood fit.

%Fig. 3
\begin{center}
\begin{figure}
\vskip -13mm
\centerline{\epsfig{file=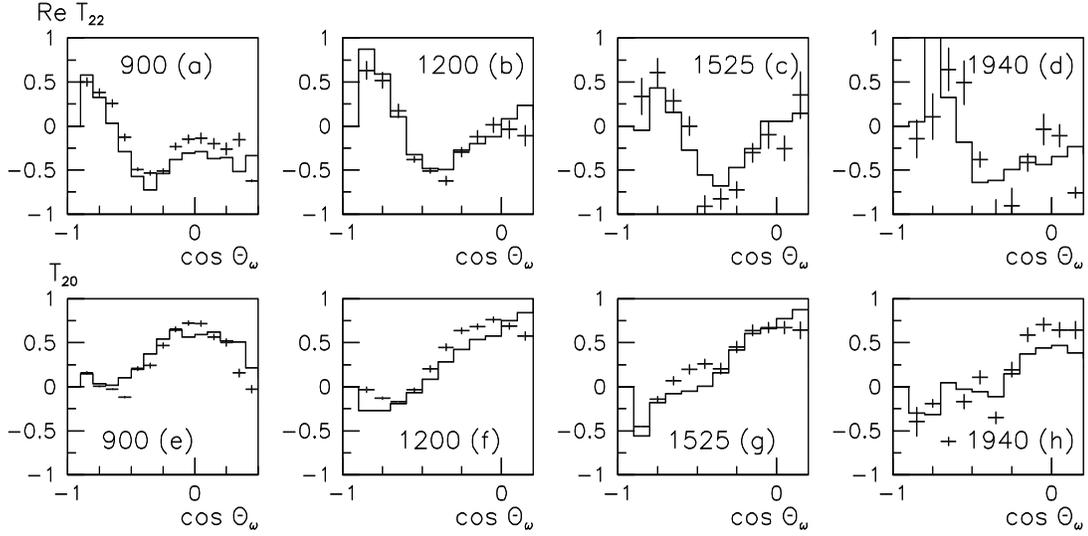,width=15cm}}
\vskip -150.25mm
\centerline{\epsfig{file=f31_wpi.ps,width=15cm}}
\
\vskip -78mm 
\caption{(a)--(d) $Re~T_{22}$ and (e)--(h) $T_{20}$ at 900, 1200,
1525 and 1940 MeV/c, compared with the partial wave fit (histograms).}
\end{figure}
\end{center}
We turn now to the partial wave analysis. This follows precisely
the lines described in earlier publications [1,2].
It has been carried out (a) at all beam momenta
separately, (b) at all beam momenta simultaneously in terms of a sum
of $s$-channel resonances.
The motivation for the latter approach comes from a parallel analysis of
extensive data from many channels with $I = 0$, $C = +1$ [8].
There, a strong $3^{--}$ resonance is required at 1985 MeV in analysis of
data on $\bar pp \to \pi^-\pi ^+$. Its interference wth the well-known
$f_4(2050)$ is observed clearly and required a resonant phase for the
$3^{--}$ partial wave.
Further $4^{++}$ and $3^{--}$ resonances are
required in the range 2250--2300 MeV.
Once these are included into the analysis of present data, relative
phases of other partial waves require resonances in other partial waves.

The parametrisation in terms of resonances introduces the important
constraint of analyticity, since Breit-Wigner  amplitudes are analytic
functions of $s$.
It smoothes out considerable fluctuations in phases
found in the analysis at individual beam momenta.
Partial wave amplitudes take the form
\begin {equation}
f = \sum _i \frac {g_i\exp(i\phi _i)B_L(q)B_{\ell}(p)}
                  {M_i^2 - s - iM_i\Gamma _i };
\end{equation}
$B_{\ell }$ and $B_L$ are standard Blatt-Weisskopf centrifugal barrier
factors for production with orbital angular momentum $\ell $ in the
$\bar pp$ channel and decay with orbital angular momentum $L$ to
$\omega \pi ^0$ (or 2-body channels in $\omega \eta \pi ^0$);
$p$ and $q$ are centre of mass momenta in $\bar pp$ and meson channels
respectively. We adopt a radius of 0.83 fm for the centrifugal barrier
radius in all partial waves up to $\ell = 3$, as determined in Ref. [8];
this radius is increased to 1.1 fm for higher partial waves.
In the Breit-Wigner denominator, the widths of the resonances $\Gamma _i$
are taken to be constant, in view of the large number of open channels.

We now describe the spectroscopic notation for
partial waves, with examples.
The initial $\bar pp$ state may have spin $s = 0$ or 1.
For singlet ($s = 0$) states, the total angular momentum is
$J = \ell$.
For triplet states $(s = 1)$, $J = \ell$ or $\ell \pm 1$.
The parity is $P = (-1)^{\ell }$ and $C = (-1)^{\ell + s}$.
In the $\omega \pi $ exit channel, the spin $s$ is 1 from the
$\omega$.
As examples, $J^{PC} = 1^{--}$ may couple to initial $\bar pp$
$^3S_1$ and $^3D_1$ partial waves having $\ell = 0 $ or 2;
the exit $\omega \pi $ channel has $L = 1$.
In the analysis, $^3S_1$ and $^3D_1$ partial waves are fitted initially 
with independent phases $\phi$ and a ratio of coupling constants
$r = g_{\ell = J+1}/g_{\ell = J - 1}$.
[In the final fit, results discussed below allow phases for 
$L =  J\pm 1$ to be constrained to the same value].
A second example is $J^{PC} = 3^{+-}$.
This couples to $\bar pp$ singlet states and decays to $\omega \pi ^0$
with $L = 2$ or 4.
These two partial waves are again fitted with independent phases $\phi$
and a ratio of coupling constants $r = g_{L = J+1}/g_{L = J-1}$.
These $r$ parameters for decays to the $\omega \pi$
channel have improved greatly in accuracy compared with earlier
work, because of the new polarisation information.

The $\omega \eta \pi ^0$ data are fitted to
sequential 2-body processes: $\omega a_2(1320)$, $\omega a_0(980)$,
$b_1(1235)\eta$ and $\omega (1650)\pi ^0$.
As regards notation,  consider $\omega a_2$.
Spins 1 of the $\omega$ and 2 for the $a_2$ combine to
total spin $s' =  1$, 2 or 3.
The initial $\bar pp$ $^1F_3$ states with $J^{PC} = 3^{+-}$ may
decay to $\omega a_2$ with $L = 1$, 3 or 5.
In practice, the lowest $L$ value is always dominant;  in
every case $L $ values above the first may be omitted because of
the strong centrifugal barrier.
It is however necessary to consider all possible values of total spin
$s'$.
Intensities of partial waves are given in Figs. 6 and 7 of Ref. 2.
There, for example, $\bar pp$ $^1F_3 \to a_2\omega$ with $L = 1$
is denoted $1F3 - a_2\omega$ $5P3$ or $7P3$ (the former with $s' =  2$
and the latter with $s' = 3$).
Again separate phases are fitted to every channel.

A flat component across the Dalitz plot
is also required in order to fit the projection of data on to
$\eta \pi$ mass.
As illustrated in Fig. 3 of Ref. [2], conspicuous narrow $a_0(980)$ and
$a_2(1320)$ peaks appear in $\eta \pi$. However the optimum fit
requires in addition some flat physics background, not originating from
$\omega a_2$, $b_1\eta$ or $\omega (1650)\pi ^0$.
This extra
component peaks for $\bar pp$ masses in the range 2100--2260 MeV and
seems to be associated with the strong $a_2(1320)\omega$ threshold at
2100 MeV.
Following decays of the $a_2$ or $a_0$, the $\eta$ and $\pi ^0$ may
rescatter from the $\omega$. Calculation of these so-called `triangle
diagrams' leads to broad components having a logarithmic variation with
$\eta \pi$ mass across the Dalitz plot; such effects are beyond the
isobar model.
The required flat component is dominantly associated with
$\bar pp$ $^3S_1$ and has the effect of broadening the $a_2\omega$
threshold. It is fitted freely as a further broad $^3S_1$ resonance of
mass 2080 MeV and width 350 MeV, but we do not claim it as a resonance.
It may be absorbed into the $^3S_1$ resonance at 2110 MeV with a modest
deterioration of log likelihood. Similar smaller effects are
needed in $^1P_1$, peaking at 2240 MeV, and in $^3D_3$ at 2260 MeV.
They are absorbed into $^1P_1$ and $^3D_3$ resonances at those masses,
as shown in Figs. 6 and 7 of Ref. [2]. These flat components hinder the
precise determination of $\bar pp$ resonance masses. Further
flat components in other partial waves have been tried, but lead to no
further improvements.

Table 1 shows that
statistics of present $\omega \eta \pi ^0$ data are quite low compared
with $\omega \pi ^0$. They play only a small role in the fit.
Intensities in the $\omega \eta \pi ^0$ partial waves have
hardly changed from those shown in Figs. 6 and 7 of Ref. [2]
and will not be repeated here; phases in this channel have however
improved appreciably because of the new polarisation information.

A fresh feature of the analysis is that parameters of resonances
visible in the $\pi ^- \pi ^+$ data are adjusted to achieve the
best fit with present data for $\omega \pi$ and $\omega \eta \pi$.
This completes a combined fit to all available data with $I = 1$, $C = -1$.
In detail, the way this is carried out is as follows.
Masses and widths of those resonances appearing in $\pi ^+\pi ^-$
(namely $\rho _1$, $\rho _3$ and $\rho _5$) are first fitted to
those data and errors are determined.
A separate fit is made to the combined $\omega \pi ^0$ and
$\omega \eta \pi ^0$ data, finding resonance masses, widths and
errors.
The weighted means of masses and widths are then formed from the
two determinations.
Using these values, final fits are made to the combined
$\omega \pi ^0$ and $\omega \eta \pi ^0$ data, and separately to
$\pi ^+\pi ^-$.
Changes at this stage are within systematic errors, which depend
on precisely how many partial waves are allowed in the fit.
In practice, the $\pi ^+\pi ^-$ data apply powerful constraints
to the $\rho _3$ fitted at 1985 MeV and to $\rho _1$ at 1970 MeV,
and a lesser constraint to $\rho _5$ at 2300 MeV.

Vector polarisation depends on the imaginary parts of interferences
between partial waves. Tensor polarisation depends on moduli squared of
amplitudes and real parts of interferences. Together they improve the
determination of phases and hence many  resonance parameters.
We find no major changes from Refs. [1] and [2], but considerable
clarification for some resonances.

Table 2 shows results of the analysis for masses and widths of
resonances. Columns 6 and 8 show changes in log likelihood when
each resonance is omitted from the fit and remaining resonances
are re-optimised. For convenience, columns 7 and 9 show
corresponding values from the earlier analyses. One sees
immediately a considerable increase in the significance of some
resonances.

\begin{table} 
\begin{center}

\caption {Resonance parameters from a combined fit to $\omega \pi ^0$,
$\omega \eta \pi ^0$ and $\pi ^- \pi ^+$, using both
$\omega \to \pi ^0 \gamma$ and $\omega \to \pi ^+\pi ^- \pi ^0$ decays.
Values in parentheses are fixed.
Values of $r$ are ratios of coupling constants
$g_{J+1}/g_{J-1}$ for coupling to $\bar pp$ or $\omega \pi ^0$.
Columns 6 and 8 show changes in
$S = $ log likelihood when each resonance is omitted from this fit and others
are re-optimised; for comparison, columns 7 and 9 show changes
observed in earlier analyses.}
\vskip 2mm
\begin{tabular}{ccccccccc} 
\hline 
$J^{PC}$ & Mass $M$ & Width $\Gamma$ & $r$ & $\phi $ & $\Delta S$ & Previous &
$\Delta S$ & Previous \\
        & (MeV)  & (MeV) &  & (deg) & $(\omega \pi )$ &
        $\Delta S (\omega \pi )$ & $(\omega \eta \pi )$ &
        $\Delta S (\omega \eta \pi )$ \\ \hline
$1^{+-}$ & $1960 \pm 35$ & $230 \pm 50$  & $0.73 \pm 0.18$ &
  -79 & 503 & 289  & 514 & 299\\
$1^{+-}$ & $2240 \pm 35$ & $320 \pm 85$ & $1.2 \pm 0.5$ &
  -136 & 62 & 79 & 542 & 363 \\
$3^{+-}$ & $2032 \pm 12$ & $117 \pm 11$ & $2.06 \pm 0.20$ &
   (0) & 3073 & 1115 & 178 & 128 \\
$3^{+-}$ & $\sim 2245$ & $320 \pm 70$ & $0.78 \pm 0.47$ &
  -161 & 264 & 346 & 187 & 185 \\
$5^{+-}$ & $(2500)$ & $\sim 370 $ & (0)  &
  -11 & 195 & 36 & - & - \\\hline
$1^{--}$ & $1970 \pm 30$ & $260 \pm 45$ & $0.70 \pm 0.23$  &
  34 & 562 & 295 & 133 & 186 \\
$1^{--}$ & $2110 \pm 35$ & $230 \pm 50$ & $-0.05 \pm 0.42$  &
  - & - &  -  & 834 & 839 \\
$1^{--}$ & $2265 \pm 40$ & $325 \pm 80$ & $-0.55 \pm 0.66$  &
  - & - & - & 313 & 430 \\
$2^{--}$ & $1940 \pm 40$  & $155 \pm 40$ & $1.30 \pm 0.38$ &
   53  & 433 & 227 & 93 & 85\\
$2^{--}$ & $2225 \pm 35$ & $335 ^{+100}_{-50}$ & $1.39 \pm 0.37$ &
   -71 & 356 & 296 & 502 & 198\\
$3^{--}$ & $1982 \pm 14$  & $188 \pm 24$ & $0.006 \pm 0.008$ &
  -171 & 314 & 64 & 138 & 61 \\
$3^{--}$ & $2260 \pm 20$ & $160 \pm 25$ & $1.6 \pm 1.0$ &
  60 & 341 & 52 & 578 & 456 \\
$4^{--}$ & $2230 \pm 25$ & $ 210 \pm 30$ & $0.37 \pm 0.05$ &
  (0) & 1254 & 1159 & 153 & 79 \\
$5^{--}$ & $2300 \pm 45$ &$ 260 \pm 75$ & (0)  &
  -38 & 473 & 33 & 133 & 78 \\\hline
\end{tabular}
\end{center}
\end{table}

We now discuss individual partial waves, beginning with the highest
$J$.
Their intensities in the $\omega \pi$ data as a function of mass are
shown in Fig. 4.
For $J^P = 5^-$, there is a very large improvement in log likelihood
for $\omega \pi$ data, from 33 to 473. [Our definition of log likelihood
is such that it changes by 0.5 for a one standard deviation change to
one fitted parameter]. There is a corresonding large improvement in
the determination of resonance mass and width, now $M = 2300 \pm 45$
MeV, $\Gamma = 260 \pm 75$ MeV.
These results are consistent within errors with those of the GAMS group
[9] for $\rho _5(2350)$, $M = 2330 \pm 35$ MeV, $\Gamma = 400
\pm 100$ MeV. We note, however, that they did not include centrifugal
barriers; these move the peak position upwards, so it is to be expected
that their resonance mass will be higher. The intensity in Fig. 4 peaks
at 2.34 GeV.

%Fig. 4
\begin{figure}
\vskip -20mm
\centerline{\epsfig{file=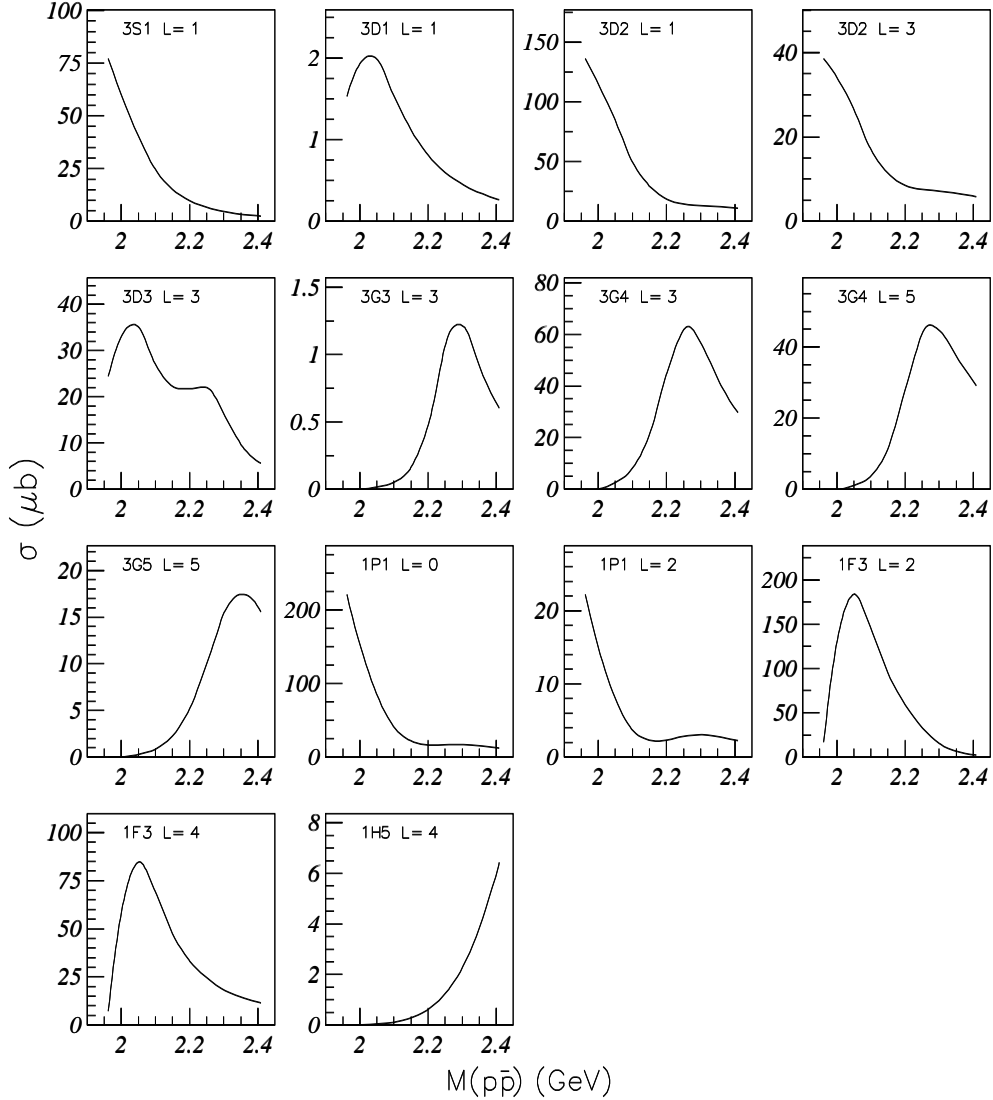,width=15cm}}
\vskip -165.35mm
\centerline{\epsfig{file=F5_WPI.PS,width=15cm}}
\vskip -6mm
\caption{Intensities of partial waves fitted to $\omega \pi ^0$ data;
$L$ is the orbital angular momentum in the decay.}
\end{figure}

For the very strong $^3G_4$ partial wave, there is very little
change from the earlier analysis of Ref. [1].
It still lies anomalously low in mass ($M = 2230 \pm 25$ MeV) compared
with other $G$ states, with mass $2230 \pm 25$ MeV. Its intensity peaks
in Fig. 4 at 2.265 GeV.

There is an improvement in the definition of the $3^{--}$ state at
2260 MeV. It is now very precisely determined, whether or not
the $^3G_3$ state expected close-by is included in the analysis.
Adding this $^3G_3$ state, log likelihood improves by 117 for the
addition of 14 parameters (including several decay channels in
$\omega \eta \pi)$. This improvement is of marginal significance.
When the $^3G_3$ mass and width are scanned, we find an optimum mass of
2270 MeV with a width of 180 MeV.
These are close to the other $3^{--}$ state, so our conclusion
is that there is no firm evidence at present for a third $3^{--}$ state.
It is likely that the upper two states will be mixed, but mixing is
expected to push them apart.

At lower mass, the $3^{-}$ state at 1982 MeV is extremely well
determined by data for $\bar pp \to \pi ^- \pi ^+$.
Data for $\omega \pi$ and $\omega \eta \pi$ give an optimum mass of
$2004 \pm 53$ MeV, $\Gamma = 270 \pm 65$ MeV.
Values quoted in Table 2 are the weighted mean for all three sets
of data.

Two states with $J^{PC} = 2^{--}$ are definitely required, but their
parameters are not accurately determined.
The upper state at 2225 MeV is reasonably well determined from
$\omega \pi$ data. It is less clear in $\omega \eta \pi$ data because
of cross-talk with strong $1^{--}$ signals in the $a_2(1320)\omega$
channel.
The lower $2^{--}$ state at 1940 MeV is close to the bottom of the
available mass range and is therefore poorly determined in mass and
particularly poorly determined in width. It is one of the two least well
determined resonances in Table 2.

Analysis of the $1^{--}$ sector is hampered by the lack
of data from a polarised target.
Our experience in Ref. [8] is that these data are vital to separate
$^3S_1$ and $^3D_1$ partial waves.
The absence of those data causes  a blurring of the large
partial waves for $1^{--}$.

There is a strong $1^{--}$ contribution required by  present data at the
lowest masses.
However, it is better determined from $\pi ^- \pi ^+$ data, where
both differential cross sections and polarisation data are
available at 100 MeV/c steps of beam momentum down to 360 MeV/c [10].
Table 2 quotes the weighted means from those data and present
data, $M = 1970 \pm 30$ MeV.
This state lies close to other $D$-states and is likely to be
the radial recurrence of the $\rho (1700)$, which is generally
believed to be the $^3D_1$ ground-state.

At higher masses, there is a distinct improvement in the mass of
the resonance at 2110 MeV.
It is visible only in $\omega \eta \pi$ data and makes negligible
contribution to $\omega \pi$.
It is likely to be the same resonance as listed by the
Particle Data Group at $2149 \pm 17$ MeV with $\Gamma = 363 \pm 50$
MeV [11].

At still higher energies, the mass scan reveals two definite
$1^{--}$ peaks at 2265 and $\sim 2400$ MeV. It is likely that these are
radial excitations of the lower two states. Unfortunately the latter is
at the top of the available mass range and is ill-determined. The state
at 2265 MeV makes negligible contribution to $\omega \pi$, but is
reasonably well determined now by $\omega \eta \pi$ data. In order to
complete the separation of $^3S_1$ and $^3D_1$ states, data from a
polarised target are needed; alternatively, diffraction dissociation of
linearly polarised photons would achieve the same separation.

%Fig. 5

\begin{figure}
\vskip -30mm
\centerline{\epsfig{file=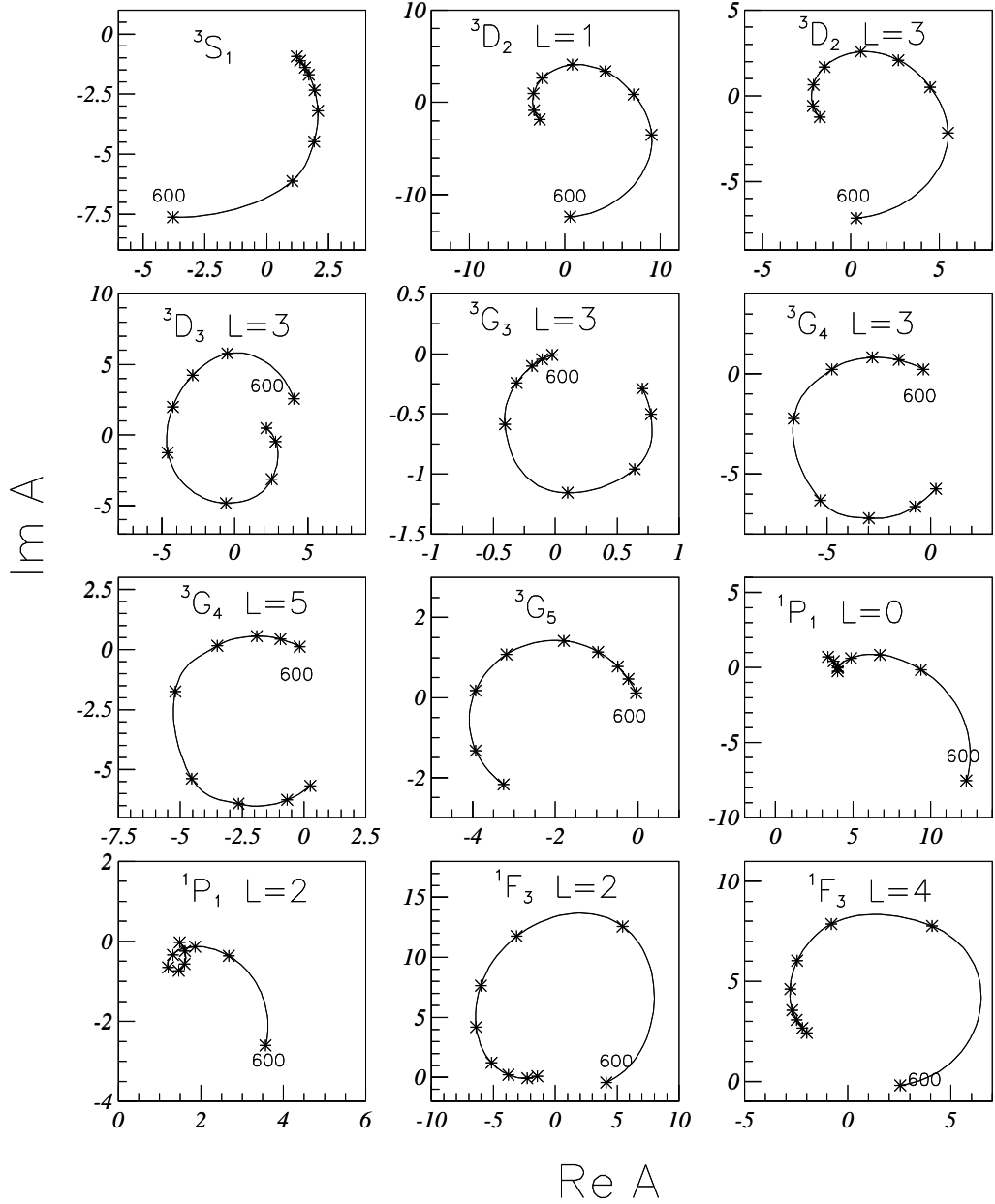,width=15cm}}
\vskip -180.3mm
\centerline{\epsfig{file=F4_WPI.PS,width=15cm}}
\caption{Argand diagrams for partial waves fitted to $\omega \pi ^0$
in the combined fit with $\omega \eta \pi ^0$ and $\pi ^+\pi ^-$.
Crosses show beam momenta 600, 900, 1050, 1200, 1350, 1525, 1642, 1800
and 1940 MeV/c; all move anti-clockwise with increasing beam momentum.
$L$ is the orbital angular momentum in the $\omega \pi$ channel.}
\end{figure}

We come now to singlet states.
The lower $3^{+-}$ state at 2032 MeV is the strongest partial wave
in $\omega \pi$ and is very well determined.
It is particularly narrow, with a well determined width of $117 \pm 11$
MeV.
In Fig. 4, the intensity in this partial wave drops to a
small value above $\sim 2220$ MeV.
A fit with $b_3(2032)$ alone fails to fit the high mass region
precisely.
The reason is that a single resonance fails to fit the $360^{\circ }$
phase variation observed on the Argand Diagram of Fig. 5.
The fit to $\omega \pi$ improves strongly with the addition of
a second $3^{+-}$ resonance at $\sim 2250$ MeV.
It also makes an improvment of 187 in log likelihood for
$\omega \eta \pi$ data.
However, in neither case is the mass determined accurately.
It is now the least well determined state in Table 2.

A feature of the new data is that they provide a considerable
improvement in the determination of the upper $1^{+-}$ state
at $2240 \pm 35$ MeV.
Its contribution to $\omega \pi$ is small, but definite.
It leads to the small structures observed in the $^1P_1$ Argand
diagrams of Fig. 5; it is better determined for $L = 2$. It is also
clearly visible now in $\omega \eta \pi$ data, where it leads to an
improvement in log likelihood of 542, an overwhelming amount.

A strong $1^{+-}$ state is definitely required at the
bottom of the available mass range, with mass $M = 1960 \pm 35$ MeV.
Its amplitude goes to zero at the $\bar pp$ threshold.
This provides an anchor point in the amplitude analysis.
However, its mass and width are somewhat sensitive to
the radius chosen for the centrifugal barrier.
Further data in the low momentum range 360--900 MeV/c
are needed to complete an accurate determination of its
parameters.

%Fig. 6
\begin{figure}
\vskip -17mm
\centerline{\epsfig{file=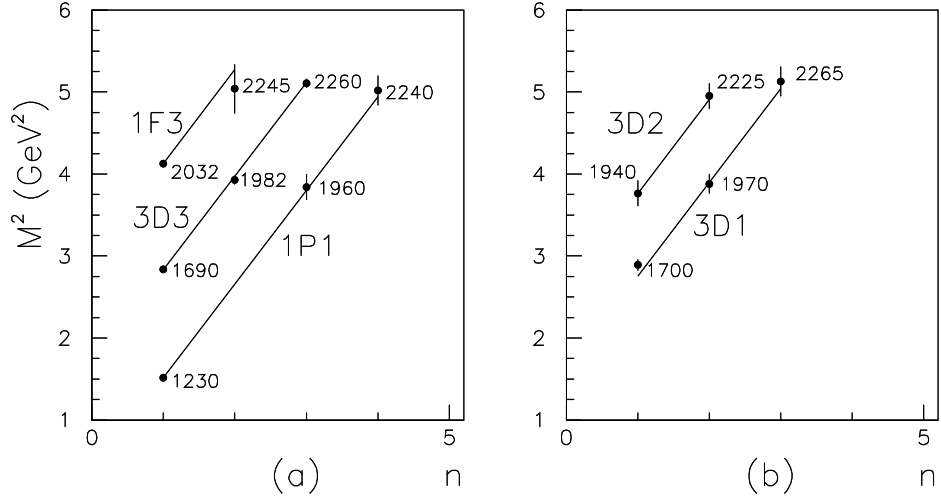,width=15cm}}
\vskip -150.3mm
\centerline{\epsfig{file=CMI1.PS,width=15cm}}
\vskip -71mm
\caption{A comparison of $M^2$ for resonances with straight-line
trajectories against radial excitation number $n$; the slope of 1.143
GeV$^2$ is taken from Ref. [8]. In (b), the $^3D_2$ trajectory is
displaced one place left in $n$, in order to resolve it from $^3D_1$.}
\end{figure}
Fig. 6 displays masses squared for resonances v.
radial excitation number.
Within errors, they follow straight-line trajectories
as in Ref. [8].
They are consistent with the identification of $q\bar q$ states
expected in this mass range.
The only missing states are $^3G_3$, expected around 2300 MeV,
and the highest $1^{--}$ state, of which there is a hint at
$\sim 2400$ MeV in present data.
It is not surprising that these states of the lowest or highest
spins are difficult to locate.

A remarkable feature of the present $\omega \pi$ data is that
the vector polarisation is close to zero everywhere.
We have checked by several methods that this is not mistaken; for example,
dropping the sign of $\vec n$ leads to a non-zero result.
As remarked above, $P_y$ depends on the imaginary part of interferences
between partial waves.
This implies that relative phases are close to 0 or $180^{\circ }$;
we now discuss the implications of this phase coherence.

Consider first a given $J^P$ such as $1^{+-}$.
This partial wave decays to $\omega \pi$ with orbital angular momentum
$L = 0$ or 2.
It is not surprising that partial waves to these final states have the same
phase, since a resonance implies multiple scattering through all coupled
channels.
Table 3 shows relative phases for several resonances.
They are consistent with zero within errors.
We therefore set them all to zero in the final analysis.

\begin{table} [htp]
\begin{center}
\begin{tabular}{ccc}
\hline
$J^{PC}$ & Mass $M$ & $\phi _{J+1} - \phi _{J - 1}$ \\
         & (MeV) &  (deg) \\\hline
$4^{--}$ & 2230 & $17.4 ^{+6.3}_{-13.9}$ \\
$3^{+-}$ & 2032 & $-1.5 \pm 8.2$ \\
$2^{--}$ & 2225 & $0.7 \pm 9.2$ \\
$2^{--}$ & 1940 & $5.4 \pm 13.8$ \\
$1^{+-}$ & 1960 & $18.4 \pm 25.5$ \\\hline
\end{tabular}
\caption {Relative phases for decays to $\omega \pi$ with $L =
J \pm 1$.}
\end{center}
\end{table}

However, vector polarisation $P_y$ can arise from interference between
singlet waves
$1^{+-}$, $3^{+-}$ and $5^{+-}$.
Likewise, it can arise from interference between triplet waves
$1^{--}$, $2^{--}$, $3^{--}$, $4^{--}$ and $5^{--}$.
It is remakable that these interferences also lead to nearly zero
polarisation.
It implies coherence between different $J^P$.
We have no full explanation for this result, but remark on some
points which may be relevant.

Firstly, it is generally assumed that different $J^P$ are
uncorrelated because of spherical symmetry.
However, the region of strong interactions is Lorentz contracted
in the collision.
Nonetheless, the time of interaction is so short that it is hard
to see how different $J^P$ could couple during the interaction.

A more important consideration is that the incident plane wave
may be expanded in the usual way in terms of Legendre functions:
\begin {equation}
e ^{ikz} = \sum _{\ell} (2\ell + 1)i^{\ell} P_{\ell}(\cos \theta ).
\end {equation}
As a result, all partial waves in the initial state are related in phase
through the factor $i^{\ell}$. Fig. 5 shows Argand diagrams for all
partial waves. Let us take as reference for triplet states the large
$^3G_4 (L = 3)$ amplitude. If one compares this by eye with diagrams
for $^3D_2$ and $^3D_3$, it is clear that there is a phase rotation of
the diagram by $\sim 180 ^{\circ}$ for $^3D_2$, but $^3D_3$ is similar
in phase to $^3G_4$. The Argand loop for $^3S_1$ is again rotated by
$\sim 180 ^{\circ }$ with respect to $^3G_4$.
Singlet and triplet states do not interfere in $d\sigma /d\Omega$,
$P_y$, $T_{20}$, $T_{21}$ and $T_{22}$.
Therefore the phase of singlet states with respect to triplet
are arbitrary in Fig. 5.
The Argand loops for $^1P_1$ and $^1F_3$ are broadly similar,
though there is a large offset in the $^1P_1$ $L = 0$ amplitude.

A similar phase coherence is reported for $\bar pp \to \pi ^- \pi ^+$
[12].
There, both $I = 0$ and $I = 1$ amplitudes are present.
The amplitudes differing by 1 in orbital angular momentum for
$\bar pp$ are $\sim 90 ^{\circ}$ out of phase, as equn. 9 suggests.

Table 2 shows phases of individual resonances in column 5. They do not
follow any simple pattern. However, they have varying masses and
widths. We propose that resonances are formed with phases which follow
on average the phase coherence implicit in equn. 9.

An analogy may be helpful from common experience.
Suppose a bottle is filled with irregular shaped objects, such as
screws.
If the bottle is shaken, the objects inside settle
to a more compact arrangement with increased order.
Returning to resonances, they have well defined masses, widths and
coupling constants making them individual.
We suggest that in the interaction, phases adjust to retain as
much coherence as possible with the incident plane wave.

In summary, the new data provide a considerable improvement in
parameters of several resonances.
This arises from more accurate polarisation information
for the $\omega$, leading to better phase determination.
There is a remarkable phase coherence between partial waves.

\section{Acknowledgement}
We thank the Crystal Barrel Collaboration for
allowing use of the data.
We acknowledge financial support from the British Particle Physics and
Astronomy Research Council (PPARC).
We wish to thank Prof. V. V. Anisovich for helpful discussions.
The St. Petersburg group wishes to acknowledge financial support
from grants RFBR 01-02-17861 and 00-15-96610 and from PPARC;
it also wishes to acknowledge support under the Integration of
the Russian Academy of Science.
%the German Bundesministerium f\"ur Bildung, Wissenschaft,
%Forschung und Technologie,
%the Schweizerischer Nationalfonds,
%the U.S.~Department of Energy, and
%the National Science Research Fund Committee of Hungary (contract
%No.~DE-FG03-87ER40323,
%DE-AC03-76SF00098, DE-FG02-87ER40315 and OTKA T023635).
%K.M.~Crowe and F.-H.~Heinsius
%acknowledge support from the A. von Humboldt Foundation.


\begin {thebibliography}{99}
\bibitem {1} A. Anisovich et al., Phys. Lett. B508 (2001) 6.
\bibitem {2} A. Anisovich et al., Phys. Lett. B513 (2001) 281.
\bibitem {2} K. Peters, Nucl. Phys. A692 (2001) 295c.
\bibitem {4} E. Aker et al., Nucl. Instr. A321 (1992) 69.
\bibitem {5} Ch. Weddigen, Nucl. Phys. A312 (1978) 330.
\bibitem {6} F. Foroughi, J. Phys. G: Nucl. Phys. 8(1982) 345.
\bibitem {7} {\it The Theory of Elementary Particles}, J. Hamilton,
(Oxford, 1959), p392.
\bibitem {8} A. Anisovich et al., Phys. Lett. B 491 (2000) 47.
\bibitem {9} D. Alde et al., Zeit. f. Phys. C66 (1995) 379.
\bibitem {10} A. Hasan et al., Nucl. Phys. B378 (1992) 3.
\bibitem {11} Particle Data Group, Euro. Phys. Journ. 15 (2000) 1.
\bibitem {12} A. Hasan and D.V. Bugg, Phys. Lett. B 334 (1994) 215.
\end {thebibliography}
\end {document}